\documentclass[conference]{IEEEtran}
\IEEEoverridecommandlockouts
\usepackage{cite}
\usepackage{amsmath,amssymb,amsfonts,nccmath}
\usepackage{algorithm,algorithmicx}
\usepackage{algpseudocode}
\usepackage{graphicx,subfigure}
\usepackage{multirow}
\usepackage{textcomp}
\usepackage{xcolor}
\usepackage{longtable}
\usepackage{comment}
\def\BibTeX{{\rm B\kern-.05em{\sc i\kern-.025em b}\kern-.08em
    T\kern-.1667em\lower.7ex\hbox{E}\kern-.125emX}}

\usepackage{array}
\newcolumntype{L}[1]{>{\raggedright\let\newline\\\arraybackslash\hspace{0pt}}m{#1}}
\newcolumntype{C}[1]{>{\centering\let\newline\\\arraybackslash\hspace{0pt}}m{#1}}
\newcolumntype{R}[1]{>{\raggedleft\let\newline\\\arraybackslash\hspace{0pt}}m{#1}}

\makeatletter
\newcommand{\linebreakand}{%
  \end{@IEEEauthorhalign}
  \hfill\mbox{}\par
  \mbox{}\hfill\begin{@IEEEauthorhalign}
}
\makeatother

\author{
  \IEEEauthorblockN{Minoo Mohebbifar}
  \IEEEauthorblockA{\textit{Department of Electrical and Computer Engineering} \\
    \textit{Tabriz University}\\
    Tabriz, Iran \\
    mohebbifar.minoo@tabriz.ac.ir}
  \and
  \IEEEauthorblockN{Mohammad Panahazari}
  \IEEEauthorblockA{\textit{Department of Electrical and Computer Engineering} \\
    \textit{Clarkson University}\\
    Potsdam, NY, USA \\
    panaham@clarkson.edu}
  \linebreakand 
  \IEEEauthorblockN{Omid Mirzapour}
  \IEEEauthorblockA{\textit{Department of Electrical and Computer Engineering} \\
    \textit{University of Utah}\\
    Salt Lake City, UT, USA \\
    omid.mirzapour@utah.edu}

}
    
\begin{document}

\title{Improved Dual-Output Step-Down Soft-Switching Current-Fed Push-Pull DC-DC Converter}

\IEEEoverridecommandlockouts


\maketitle

\IEEEpubidadjcol

\begin{abstract}
Multi-port DC-DC converters are gaining more significance in modern power system environments by enabling the connection of multiple renewable energy sources, so the efficient operation of these converters is paramount. Soft switching methods increase efficiency in DC-DC converters and increase the reliability and lifespan of devices by relieving stress on components. This paper proposes a method for soft-switching of a dual-output step-down current-fed full-bridge push-pull DC-DC converter. The converter enables two independent outputs to supply different loads. The topology achieves zero-current switching on the primary side and zero-voltage switching on the secondary side, eliminating the need for active-clamp circuits and passive snubbers to absorb surge voltage. This reduces switching losses and lower voltage and current stresses on power electronic devices. The paper thoroughly investigates the proposed converter's operation principle, control strategy, and characteristics. Equations for the voltage and current of all components are derived, and the conditions for achieving soft switching are calculated. Simulation results in EMTDC/PSCAD software validate the accuracy of the proposed method.

\end{abstract}


\section{Introduction}
The rapid deployment of distributed energy resources (DERs) in distribution grids has spurred great interest in the DER management system (DERMS) for both distribution system operators (DSOs) and transmission system operators (TSOs) \cite{mirzapour2023transmission}. DERs span a wide range of technologies, including rooftop solar photovoltaic (PV) panels, small wind turbines, residential battery energy storage systems, electric vehicles (EV), EV charging stations (EVCS), and controllable loads~\cite{nematirad2022solar}. The deployment of DERs brings several benefits to the power systems, including increased reliability, resilience and cost saving~\cite{mirzapour2022multidimensional}. However, most DER technologies are DC and require an efficient interface with the power system. The increasing adoption of DC systems in various industries highlights the importance of DC-DC converters. The significant role of DC technologies in electricity generation, storage, and management further highlights this demand for improvements in DC-DC converters~\cite{salehi2021predictive}. Efficient voltage regulation and power delivery are crucial for reliable operations, driving the need for advancements in DC-DC converter technology \cite{feizi2021mirrored}. The increasing need for flexibility in modern power systems and the adaptability of DC-DC converters makes them suitable for many systems, including DC and hybrid AC/DC systems~\cite{rezaei2022hybrid}.  Minimizing switching losses is a key challenge, which can be addressed through effective soft-switching methods \cite{ganjavi2018novel}. Soft-switching techniques enhance converter performance and efficiency, facilitating the integration of DC systems in diverse applications.

Multiport DC-DC converters offer a compact and cost-effective solution compared to separate converters \cite{rasouli2022lyapunov}. Multiport DC-DC converters efficiently manage distributed systems, especially with multiple energy sources and loads, including PV panels, fuel cells, battery storage, and electric vehicles. Multiport DC-DC converters enable efficient energy flow and power exchange among these energy sources and loads~\cite{siratultra}. The superior performance of multi-port converters over conventional converters in highly renewable integrated systems further indicates their merit as the new alternative. However, the high voltage stress on switches poses a significant challenge, impacting efficiency and cost due to the limited availability of high-voltage switches with high forward voltage drop and ON-state resistance~\cite{notash2018multi}. The high voltage switching further adds to switching loss from transition, reducing the converter's efficiency and creating challenges for the thermal management of the converter~\cite{mirzapour2018evaluating}. Finally, the high voltage switching creates voltage and current transients, increasing the electromagnetic interference and potentially damaging the converter~\cite{noruzi2020variable,norouzi2023stability}. Phase-shifted modulation has also been proposed in the literature, which can reduce switching losses by spreading switching losses over a wider range by creating phase shifts among switches~\cite{liu2017dual}. This method, however, requires additional switches and more complex control schemes and suffers from issues including increased electromagnetic interference and limited load range~\cite{sirat2020new}.

Previous studies explored multi-input/multi-output converters and active-clamp circuits~\cite{ghaderloo2023high}. These approaches utilized coupled inductors and voltage multiplier cells but suffered from increased cost, volume, and power losses. Another approach employed switched capacitors and coupled inductors for high voltage conversion ratios and reduced voltage stress \cite{he2019soft, cheshmdehmam2018soft}. In a recent work \cite{kothapalli2022zvs}, a dual coupled inductor-based flyback energy conversion circuit achieved high voltage step-up/down ratios and efficiency. An active switch-based capacitor multiplier cell reduced voltage stresses and minimized clamp device count. Quasi-resonance enabled ZVS and ZCS across a wide load range, enhancing efficiency.

Building upon previous research \cite{previousWork}, this paper proposes an improved dual-output step-down soft-switching push-pull DC-DC converter with reduced device count. Soft-switching techniques are applied to all devices to minimize switching losses~\cite{kazemtarghi2023asymmetric}. The converter delivers two independent output voltages to supply two loads, achieving zero-current switching on the primary side and zero-voltage switching on the secondary side. The voltage and current equations for all intervals are derived, and the converter's performance is validated through simulations using EMTDC/PSCAD software.

\section{Proposed dc-dc converter}
\begin{figure}[t]
\centering
\includegraphics[width=8.6cm]{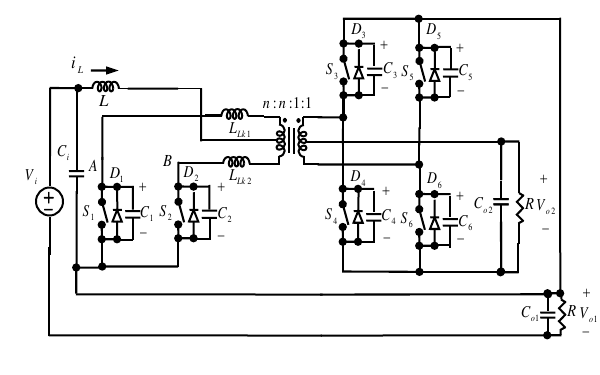}
\vspace{-0.5cm}
\caption{The Proposed dual-output step-down push-pull dc-dc converter }
\label{proposedCircuit}
\vspace{-0.5cm}
\end{figure}

\vspace{-1pt}
Fig. \ref{proposedCircuit} illustrates the power circuit of the dual-output step-down push-pull DC-DC converter. The converter utilizes a DC input power supply denoted as $V_i$ The internal diodes of the switches $S_1, ..., S_6$ are represented by the diodes $D_1, ..., D_6$. Additionally, the parasitic capacitors associated with switches $S_1, ..., S_6$ are denoted as capacitors $C_1, ..., C_6$ , respectively. An input inductor (L) is employed to reduce the input current ripple. Soft-switching capability is achieved by utilizing the leakage inductance $L_{LK}$ and parallel parasitic capacitors in conjunction with the switches. The output filter capacitors $C_{o1}$ and $C_{o2}$, serve as capacitors for the respective outputs. The transformer acts as a communication bridge between the primary and secondary sides of the converter. Furthermore, the four-winding transformer enables the provision of two outputs in this converter. Notably, the dual-output functionality is achieved by using a three-winding transformer on the secondary side.

In the proposed converter, the primary-side switches ($S_1$, $S_2$) are turned on first, followed by the secondary-side switch pairs ($S_3$, $S_6$), while the other pairs ($S_4$, $S_5$) are turned off sequentially. This creates a reflected output voltage $2nV_{o2}$ across the transformer primary, increasing the current through the switch pairs. Conversely, a reflected output voltage $-2nV_{o2}$ across the transformer primary decreases the current through the switch pairs. This achieves ZCS and allows the inverse current to flow through the body diodes of the switch pairs. As a result, the gate signals are eliminated due to ZCS on the primary-side devices, and the capacitance of the commutated devices begins to charge.

The analysis of the converter assumes the following:
\begin{enumerate}
\item All elements are ideal.
\item The input inductance is large enough to maintain a constant current.
\item The inductor represents the leakage inductance of the transformer.
\item The magnetizing inductance of the transformer is infinitely large.
\end{enumerate}

The gate signals of the switch pairs in the secondary side exhibit similarity. Specifically, the gate signal of the first-bridge switches ($S_1$ and $S_2$) and the gate signal of the second-bridge switches operate with a phase shift of 180 degrees, resulting in an overlapping gate signal for the first-bridge switch pairs. It is important to note that the duty cycle of the first-bridge switches is greater than 50\%, while the duty cycle of the second-bridge switches is less than 50\%. The steady-state waveforms for all intervals are depicted in Fig.\ref{waveforms} Furthermore, the second period includes a conducting symmetrical component in its half cycle, similar to the preceding half cycle.

\subsection{Operating Modes}
There are eight Operating modes of the proposed converter:
\begin{enumerate}
\item{Mode 1 ($t_0\leq$t$<t_1$)} : During this interval, $S_1$ on the primary side is switched on, while the anti-parallel body diodes $D_1$ and $D_2$ on the secondary side are conducting. The transfer of input power to the load is achieved through a high-frequency transformer. The voltage across the turned-off switches on the secondary side ($S_3$ and $S_4$) is $V_{o1}=2V_{o2}$, and the voltage across the turned-off switch on the primary side is equal to the reflected output voltage from the secondary side ($2nV_0$). The current values of different components are $i_{LK_1} = 0, i_{LK_2}=I_L, i_{s_1}=i_{s_4}=i_{s_5}=0, i_{s_2}=I_L, i_{D_3} = i_{D_6} = - \frac{nI_L}{2}.$ The voltage across the switches are $v_{S_1}=2nV_{o2}, v_{S_4}=v_{S_5}=v_{o1}, v_{S_2}=v_{S_3}=v_{S_6}=0.$.
Based on the given equations $i_{LK_1}$, $i_{LK_2}$, $I_L$ and $V_{o1}$ represent the currents through the leakage inductances, current of the input inductor, and output voltages, respectively.
\item {Mode 2 ($t_1\leq$t$<t_2$)}: At the beginning of this interval, the primary-side switches  $S_1$ and  $S_4$ are turned on. During this operating mode, the snubber capacitors  $C_1$ and  $C_1$ are discharged. Towards the end of this operating mode, the voltage value of the capacitors linearly decreases to zero. The current and voltages of switches will be similar to the first operating mode. The voltage of capacitor  $C_1$ will be as follows:\begin{equation}
\begin{split}
    v_{C_1} =& v_{C_1}|_{t=t_1} + \frac{1}{C_1} \int_{t_1}^{t} i_{C_1} \,dt \\
            =&2nV_{o2} + \int_{t_1}^{t} i_{LK1} = 2nV_{o2}
\end{split}
\label{eq:mode_2_1}
\end{equation}

\item {Mode 3 ($t_2\leq$t$<t_3$))}: During this interval, both primary-side switches are conducting. The reflected positive output voltage across leakage inductance $L_{LK1}$ causes its current to increase linearly, while the reflected negative output voltage across leakage inductance  $L_{LK2}$ causes its current to decrease linearly. As a result, switch $S_1$ starts conducting with zero current, reducing the current stress on the switches. The current values of the elements are determined as follows: 
\begin{subequations}
    \begin{align}
    \begin{split}
          i_{LK1} = i_{S_1} &= i_{LK1}|_{t=t_2} + \frac{1}{L_{LK_T}} \int_{t_2}^{t} 2nV_{o2} \,dt \\
                        &= \frac{2nV_{o2}}{L_{LK_T}}(t-t_2)     
    \end{split} \\
    \begin{split}
          i_{LK2} = i_{S_2} &= i_{LK2}|_{t=t_2} + \frac{1}{L_{LK_T}} \int_{t_2}^{t} -2nV_{o2} \,dt \\
                        &=I_L - \frac{2nV_{o2}}{L_{LK_T}}(t-t_2)     
    \end{split}  
    \end{align}
\label{eq:mode_2_1}
\end{subequations}
By considering the equation of the transformer turn ratio, the current of the anti-parallel body diodes can be calculated as follows:
\begin{equation}
    i_{D_1} = i_{D_6} = n(i_{LK2}-i_{LK1})= nI_L-\frac{4n^2V_{o2}}{L_{LK_T}}(t-t_2)
\end{equation}
In the above equation, $n$ represents the transformer turn ratio. Based on calculations, the total leakage inductance will be : $L_{LK1}+L_{LK2} = L_{LK_T}$.  Simultaneously with the conducting anti-parallel diodes $D_6$ and $D_3$, switches $S_6$ and $S_3$ have the capability to receive pulses to turn on with ZVS. Towards the end of this operating mode, the diodes $D_6$ and $D_3$ naturally turn off. The polarity of the transformer voltage changes due to the conducting secondary-side switches $S_6$ and $S_3$. At the end of this interval, the current values of different components are $i_{s_1}=i_{s_2}=\frac{I_L}{2}, i_{D_3} = i_{D_6} = 0,  v_{S_1}=v_{S_2}=v_{S_3}=v_{S_6}=0,  v_{S_4}=v_{S_5}=2V_{o2}$.

\item{Mode 4 ($t_3\leq$t$<t_4$)}: In this interval, secondary-side switches $S_6$ and $S_3$ are turned on in the ZVS state, resulting in reduced voltage stress on the switches. The current through the turned-on switch $S_1$ starts increasing with the same slope as the third operating mode, while the current through the turned-on switch $S_2$ starts decreasing with the same slope as the third operating mode. By the end of this mode, the current through switch $S_2$ decreases to zero, achieving ZCS state. Additionally, the current through switch $S_1$ reaches the $I_L$ value. At the end of this mode, the current values and voltage values of various devices are:$    i_{s_1}=i_{LK1}=I_L, i_{S_2} = i_{S_4} = i_{S_5}=0,i_{S_3} = i_{S_6}=\frac{nI_L}{2},
    v_{S_1}=v_{S_2}=v_{S_3}=v_{S_6}=0,  v_{S_4}=v_{S_5}=V_{o1}$.

\begin{figure}[t]
\centering
\includegraphics[width=8cm]{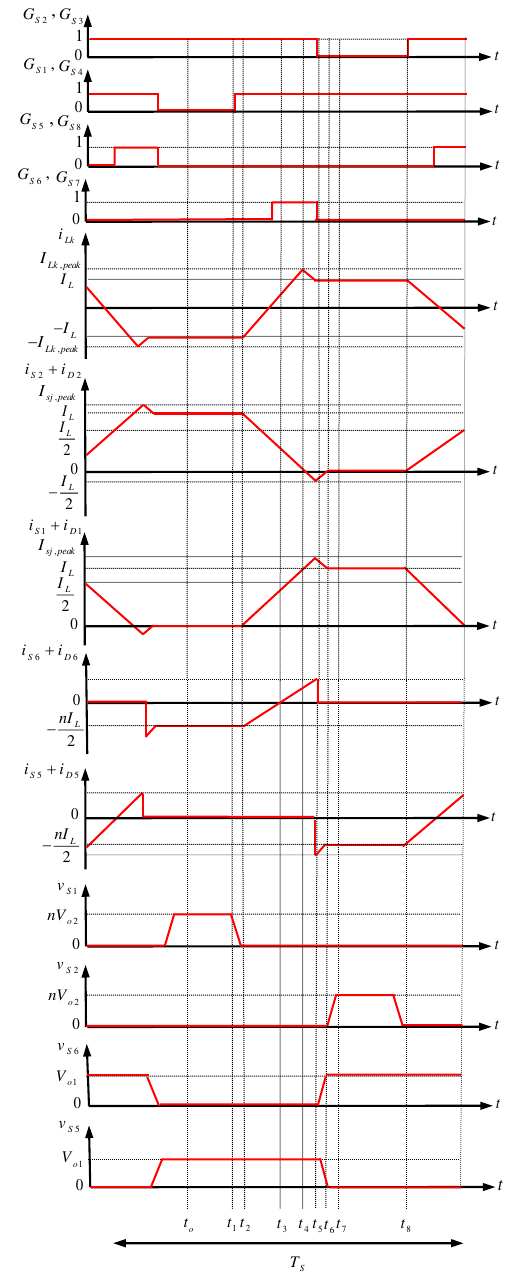}
\caption{The waveforms of converter with proposed method}
\label{waveforms}
\end{figure}

\item{Mode 5 ($t_4\leq$t$<t_5$)}: In this interval, the current through the turned-on switch $S_1$ and leakage inductance $L_{LK1}$ increases with the same fourth operating mode slope. Similarly, the current through the turned-on switch $S_2$ and leakage inductance $L_{LK2}$ decrease with the same third operating mode slope. Since the switches $S_2$ were conducting in the previous mode and turned off in the ZCS at the end of the fourth operating mode, the anti-parallel diode $D_2$ starts conducting in this mode. As a result, the voltage across this diode and the switches becomes zero for an extended interval. Additionally, at the end of this operating mode, secondary-side switches $S_6$ and $S_7$ are turned off. By using the inductor current equation for $L_{LK1}$ and $L_{LK2}$, the current through the devices can be calculated as follows:
\begin{subequations}
    \begin{align}
    \begin{split}
             i_{S_1} = i_{LK1}=I_L + \frac{2nV_{o2}}{L_{LK_T}}(t-t_4)   
    \end{split} \\
    \begin{split}
                 i_{D_2} =\frac{2nV_{o2}}{L_{LK_T}}(t-t_4)  
    \end{split}  
    \end{align}
\label{eq:mode_4}
\end{subequations}
At the end of fifth operating mode, based on transformer turn ratio, the current value and the voltage values of switches are:$
    \frac{i_1}{i_2}=\frac{1}{n}, 
    i_{s_3}=i_{s_6}=\frac{n}{2}(i_{LK1}-i_{LK2})
    =\frac{nI_L}{2}+\frac{2n^{2}V_{o2}}{LK_T}(t-t_4),
    v_{S_1}=v_{S_2}=v_{S_3}=v_{S_4}=v_{S_6} = v_{S_7} = 0,  v_{S_5}=v_{S_8}=-2V_{o2}=-V_{o1}$.
In these equations $i_1$  and $i_2$ are the current of primary-side and secondary-side of transformer. 

\item{Mode 6 ($t_5\leq$t$<t_6$)}:During this mode, when the secondary-side switches $S_6$ and $S_3$ are turned off, the anti-parallel diodes $D_5$ and $D_4$ start conducting. The voltage polarity across the transformer changes as the secondary side's current direction is reversed. As a result, the current through switch S1 and diode D2 is reduced. By using the inductor current equations $L_{LK1}$ and $L_{LK2}$, the current through the devices can be determined as follows:
\begin{subequations}
    \begin{align}
    \begin{split}
             i_{S_1} = i_{LK1} &= I_{S_i}|_{t=t_5} + \frac{1}{L_{LK_T}} \int_{t_5}^{t} -2nV_{o2} \,dt \\
                        &=I_{LK1,peak} - \frac{2nV_{o2}}{L_{LK_T}}(t-t_5)   
    \end{split} \\
    \begin{split}
                 i_{D_2} = i_{LK2} &= I_{S_{i, peak}}|_{t=t_5} + \frac{1}{L_{LK_T}} \int_{t_5}^{t} -2nV_{o2} \,dt \\
                        &=I_{D_2,peak} - \frac{2nV_{o2}}{L_{LK_T}}(t-t_5)  
    \end{split}  
    \end{align}
\label{eq:mode_4}
\end{subequations}
In the above equation, $I_{LK1}$ represents the maximum current through the leakage inductance, and $I_{D_2}$ represents the maximum current of diode $D_2$. Additionally, $S_i$ refers to the primary-side switches, where i can be either 1 or 2. The currents through diodes $D_4$ and $D_5$ are as follows:
\begin{equation}
\begin{split}
    &i_{D_4}=i_{D_5}=\frac{n}{2}(I_{LK1} - I_{LK2}) \\
    &=\frac{n(I_{LK1,peak}-I_{D2,peak})}{2}-\frac{2n^2V_{o2}}{L_{LK_T}}(t-t_5) 
\end{split}
\label{eq:mode_3}
\end{equation}
At the end of the sixth operating mode, the current through diode $D_2$ decreases to zero and turns off naturally. The current through switches $S_1$ and the primary-side transformer reach the value of $I_L$. The current and voltage values of the devices at this point are $i_{s_1}=i_{LK1}=I_L, i_{D_2} = i_{S_3} = i_{S_6}=0,
    i_{D_4}= i_{D_5}=\frac{nI_L}{2},  v_{S_1}=v_{S_2}=v_{S_3}=v_{S_4}=v_{S_5}=v_{S_8}=0,v_{S_3}=v_{S_6}=V_{o1}$.

\item{Mode 7 ($t_6\leq$t$<t_7$)}:In this short interval, the capacitors C2 are charged to a value of $2nV_{o_2}$. Additionally, the switch S2 is turned off. At the end of the seventh operating mode, the current and voltage values of the components are $   i_{LK_1}=i_{S_1}=I_L, i_{D_2} = i_{S_3} = i_{S_6}=0,
    i_{D_4}= i_{D_5}=\frac{nI_L}{2}, v_{S_1}=v_{S_4}=v_{S_5}=0,v_{S_3}=v_{S_6}=V_{o1},v_{S_2}=2nV_{o2}$.
\item{Mode 8 ($t_7\leq$t$<t_8$)}:In this interval, the switch $S_1$ is conducting. The current value of the primary-side switch $S_1$ is equal to the $I_L$ value. The value of the anti-parallel diodes of the secondary-side switches $D_4$ and $D_5$ is equal to $\frac{nI_L}{2}$. At the end of the eighth operating mode, the current and voltage values of the components are $
    i_{LK_1}=i_{S_1}=I_L, i_{D_2} = i_{S_3} = i_{S_6}=0,
    i_{D_4}= i_{D_5}=\frac{nI_L}{2}, 
    v_{S_1}=v_{S_4}=v_{S_5}=0,v_{S_3}=v_{S_6}=V_{o1},
    v_{S_2}=nV_{o1}=2nV_{o2}$.
\end{enumerate}

\section{analysis and design considerations of the proposed converter}

\subsection{Voltage Gain Calculation}
The maximum voltage across primary switches according to transformer turn ratio equation is calculated as follow:
\begin{equation}
\begin{split}
    \frac{v1}{v2} = \frac{n}{1}=\frac{\frac{v_{S_i,peak}}{2}}{V_{o2}}, 
\end{split}
\label{eq:voltageGain}
\end{equation}
In the above equation, $S_i$ is primary-side switches where $i=1,2$. By applying KVL in transformer primary-side and considering (\ref{eq:voltageGain}) the voltage gain equation is resulted as follow:
\begin{equation}
\begin{split}
    V_{o2} = \frac{V_i}{2(n(1-D)+1)}.
\end{split}
\label{voltageGain2}
\end{equation}
It should be noted that $D$ is duty cycle of the primary switches.
\subsection{Leakage Inductance Calculation}
Leakage inductance or series inductance ($L_{LK1}$, $L_{LK2}$) in a transformer can be obtained by applying KVL to the transformer's primary side and using the equation relating current and voltage of the leakage inductance:
\begin{equation}
\begin{split}
    L_{LK1} = L_{LK2} = \frac{nV_{o2}(D-0.5)}{2I_Lf_s}.
\end{split}
\end{equation}
\subsection{Input Inductance Calculation}
By applying KVL to the primary side of the transformer and considering the equation related to the voltage of the input inductor, the value of the input inductor can be calculated:
\begin{equation}
\begin{split}
    L = \frac{3n(D-0.5)}{8\Delta I_if_s(n(1-D)+1)}V_i.
\end{split}
\end{equation}
These equations are valid when the diode conduction time (sixth operating mode) is very short and can be neglected. Additionally, the primary switches are in a ZCS state without considering an increase in peak current. However, in cases where the diodes have a longer conduction time for certain loads, the output voltage is maintained higher than the voltage value of the buck converter. In such scenarios, equation (\ref{voltageGain2}) varies as follows:
\begin{equation}
\begin{split}
    V_{o2} = \frac{V_i}{2(n(1-D'-D")+1)}.
\end{split}
\label{voltage3}
\end{equation}
In the above equation, the additional time of diode conduction is denoted by $D'$, which is incorporated into the voltage gain of the converter using the proposed method. Taking into account these considerations, the equation for determining the leakage inductance is as follows:
\begin{equation}
\begin{split}
    L_{LK_T} = \frac{nV{_{o2}}(D+D"-0.5)}{I_Lf_s}.
\end{split}
\label{leakage}
\end{equation}
By simplifying (\ref{leakage}), it is resulted that:
\begin{equation}
\begin{split}
    D" = D-0.5 - \frac{I_LL_{LK}f_s}{nV_{o2}}.
\end{split}
\label{leakage}
\end{equation}
In full load conditions, equation (\ref{voltage3}) simplifies to equation (\ref{voltageGain2}) since the $D'$ is equal to zero.

\section{Simulation and Results}
The proposed converter was simulated using the PSCAD/EMTDC software program to verify its correct operation. The simulation results were obtained using the parameters listed in Table \ref{parameters}. Furthermore, soft switching was implemented for all of the switches in the simulation. The output voltages are equal to $\frac{V_{o1}}{V_i}=\frac{15.15}{48},\frac{V_{02}}{V_i}=\frac{7.54}{48}$. 

The gate signals for the switches of this converter are generated using the pulse width modulation (PWM) control method. In this control method, the duty cycle is determined by considering two factors: the carrier waveform ($A_c$) and the reference waveform ($A_r$). If the value of $A_r$ is greater than $A_c$, the comparator output will be set to 1. As a result, a firing pulse is generated for switch $S_1$, causing it to turn on. Conversely, if the comparator output is 0, switch $S_1$ is turned off. The gate signal for the other switch is generated with a shift phase relative to the firing pulse produced for switch $S_1$.

\begin{figure}[t]
\centering
\includegraphics[width=3.2in]{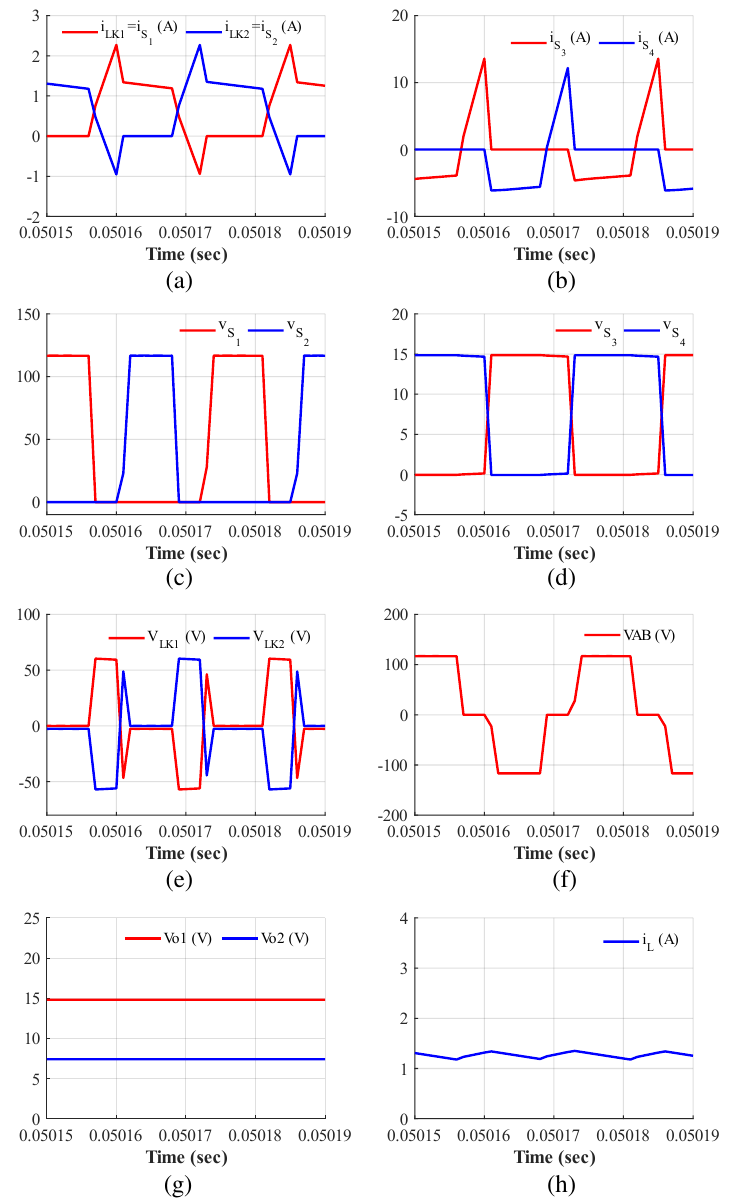}
\caption{Simulation results}
\label{results}
\end{figure}

The accuracy of the equations and waveforms is demonstrated in Fig.\ref{results}. Fig.\ref{results}(a)  and Fig.\ref{results}(c) illustrate that the switches $S_1$ and $S_2$ are turned off at zero-current switching. In this mode, the current flowing through these switches becomes zero when the voltage across them reaches an equal value. As depicted in the figures, this reduces the current stress on the primary-side switches. Fig.\ref{results}(b) and Fig.\ref{results}(d) demonstrate that the switches $S_4$, $S_5$, $S_3$, and $S_6$ are turned on at zero-voltage switching. In this mode, the voltage across the switches is initially zero when they start conducting. It should be noted that the current flowing through the primary-side switches has a lower value compared to the secondary side. Additionally, the voltage and current waveforms of the primary-side switch pair ($S_1$, $S_2$) and the secondary-side switch pairs ($S_4$, $S_5$) and ($S_3$, $S_6$) operate with a phase shift of 180 degrees. The current through the input inductor is depicted in Fig.\ref{results}(h). Based on these figures, the frequency of the input current ripple is twice the switching frequency. Table \ref{comparison} presents a comparison between the values obtained from mathematical analysis and the results from simulation. The calculated values demonstrate the accuracy of the equations derived through the proposed method in the mathematical analysis of the converter.

\begin{table}[t]
\centering
\caption{Components Specification.}
\label{parameters}
\begin{tabular}[c]{ |c|c|c|c| } 
\hline
Symbol & Parameter & Value \\
\hline
$V_i$ & Input voltage source & 48$V$ \\ 
$L$ & Input Inductor & 1109$\mu H$ \\ 
$L_{LK1}$ & Leakage inductance & 1109$\mu H$ \\ 
$L_{LK2}$ & Leakage inductance & 1109$\mu H$ \\ 
$C_i$ & Input capacitor & 4700$\mu F$ \\ 
$C_o$ & Output capacitor & 220$\mu F$ \\ 
$\frac{n}{1}$ & Transformer turn ratio & $\frac{8}{1}$ \\ 
$D$ & Duty cycle & 0.67 \\ 
$R$ & Output load & 4.5$\Omega$ \\
$f_s$ & Switching frequency & 40 $kHz$\\
\hline
\end{tabular}  
\vspace{-0.2cm}
\end{table}

\begin{table}[t]
\centering
\caption{ Comparing of obtained values of mathematical analysis and simulation.}
\label{comparison}
\begin{tabular}[c]{ |C{1.5cm}|C{3cm}|C{2.7cm}| } 
\hline
Parameter & Obtained value of mathematical analysis & Obtained value of simulation \\
\hline
$V_{o1}$ & 13.78 $V$ & 15.15$V$ \\ 
$V_{o2}$ & 6.89 $V$ & 7.54 $V$ \\ 
$V_{P_{S_j}}$ & 55.12 $V$ & 57.3 $V$ \\ 
$I_{LK2, peak}$ & 1.06 $A$ & 1.17 $A$ \\ 
$I_{S_j, peak}$ &  0.89 $A$ & 0.91 $A$ \\ 
\hline
\end{tabular}  
\vspace{-0.3cm}
\end{table}

\vspace{-0.1cm}
\section{Conclusion}
This paper presents a novel method for a dual-output step-down soft-switching current-fed full-bridge DC-DC converter. The proposed method enables effective voltage control while achieving soft-switching operation. It utilizes the secondary-modulation method for zero-current switching on the primary side and zero-voltage switching on the secondary side. Equations are derived for calculating the voltage and current of the converter's elements in all operating modes. This approach offers benefits such as reduced switching losses and minimized voltage/current stress. Importantly, the converter achieves zero-current commutation, eliminating the need for additional circuits. Simulation results confirm the precision of the proposed method using the EMTDC/PSCAD software.

\bibliographystyle{IEEEtran}
\bibliography{IEEEabrv,Reference}

\end{document}